\documentclass{ifacconf}

\usepackage{graphicx}
\usepackage{booktabs}
\usepackage{amsmath}
\usepackage{amssymb}
\usepackage{bm}
\usepackage{natbib}       
\usepackage{xcolor}
\usepackage{dsfont}
\usepackage{tikz}
\usepackage{float}
\usetikzlibrary{positioning,shapes.geometric,arrows.meta}
\usepackage{epstopdf}
\AppendGraphicsExtensions{.epsc}
\usepackage{caption}

\newtheorem{asm}{\textbf{Assumption}}

\begin{document}
\begin{frontmatter}

\title{Autopilot System for Depth and Pitch Control in Underwater Vehicles: Navigating Near-Surface Waves and Disturbances\thanksref{footnoteinfo}} 

\thanks[footnoteinfo]{This research was supported by the Office of Naval Research (grants N000142212634 and N000142112091).}

\author[First]{Vladimir Petrov} 
\author[First]{Gage MacLin} 
\author[First]{Venanzio Cichella} 

\address[First]{University of Iowa, Mechanical Engineering, Iowa City, IA 52242 USA (e-mail: \{vladimir-petrov, venanzio-cichella\}@uiowa.edu).}

\begin{abstract}                
This paper introduces a framework for depth and pitch control of underwater vehicles in near-surface wave conditions. By effectively managing tail, sail plane angles and hover tank operations utilizing a Linear Quadratic Regulator controller and \(\mathcal{L}_1\) Adaptive Autopilot augmentation, the system ensures balanced control input distribution and significantly attenuates wave disturbances. This development in underwater vehicle control systems offers potential for improved functionality across a range of marine applications. The proposed framework is demonstrated to be robust in wave conditions, enabling more precise navigation and improved safety in operational scenarios. The effectiveness of this control strategy is validated through extensive simulations using the Joubert BB2 model. 
\end{abstract}

\begin{keyword}
Joubert BB2, Near Surface Depth Keeping, \(\mathcal{L}_1\) Adaptation, LQR
\end{keyword}

\end{frontmatter}

\section{Introduction}

The advancing technological capabilities of Autonomous Underwater Vehicles (AUVs) have allowed for their application in many different areas including marine geoscience, environmental monitoring (\cite{Petillo2014Exploiting}) and national security (\cite{McNelly2022Evaluating}). The advanced applications often necessitate AUV operation in hazardous, high sea-state conditions. To ensure safety, the AUV must maintain depth and attitude with minimal actuator wear, necessitating sophisticated controllers that counter disturbances while optimizing control effort.

Depth and pitch control, a fundamental aspect of AUV operation, is critical for performing tasks at various sea levels. Research in this area, including studies by \cite{medvedev2017depth, hong2010depth, chen2014depth}, has advanced depth and pitch control techniques but has also highlighted ongoing challenges in addressing the nonlinear and time-varying nature of underwater environments. While considerable advances have been made in depth control for AUVs, existing methods struggle with the challenges in the dynamic, nonlinear, underwater environment, particularly in the presence of surface waves and low-frequency disturbances. Addressing these challenges is the primary focus of this research.

Recent progress in addressing these challenges, as illustrated by the research conducted in \cite{Steenson2012Experimental, Wang2018Coordinated, Ajaweed2023Submarine}, has been primarily focused on the implementation of a wide spectrum of control architectures. These methodologies include robust adaptive control, model predictive control, and sliding mode control. However, these studies do not address the unique disturbances that occur near the water surface, such as wave effects and suction phenomena. In contrast, \cite{DANTAS2012319} focused on addressing the issue of wave disturbances in the motion of AUVs. They proposed a controller structure based on the LQG/LTR methodology with a wave filter. While this study provides a control architecture capable of managing wave disturbances, it does not encompass the challenges posed by low surface suction anomalies that AUVs frequently encounter. Incorporating an adaptive augmentation method could serve as a valuable extension to current control strategies, enhancing the robustness and adaptability of AUVs in dealing with these specific anomalies, especially in shallow water environments.

In this research paper, we introduce a control architecture designed to effectively eliminate waves from the actuator and mitigate steady state errors arising from low-frequency disturbances and uncertainties. This architecture features two key elements: an autopilot, i.e., a refined Linear Quadratic Regulator (LQR) controller enhanced with filtering capabilities, and an advanced \(\mathcal{L}_1\) autopilot augmentation system. The LQR controller, integrated with a filtering mechanism, is tasked with regulating the tail, sailplane angles, and hover tank, providing wave disturbance rejection.  This ensures a balanced distribution between these control inputs, thereby diminishing the strain experienced by individual components in challenging environments, such as near surfaces with the presence of waves. The \(\mathcal{L}_1\) adaptive controller is designed to counteract steady-state errors caused by low-frequency disturbances and model uncertainties, including suction forces. The control architecture is validated using the Joubert BB2 reduced order model, as introduced in \cite{martin2022reduced}, which incorporates uncertainties due to unmodeled dynamics and realistic near surface excitation.

This paper is structured as follows: Section 2 discusses the Joubert BB2 vehicle's control dynamics and the problem at hand. Section 3 presents the design of the LQR Autopilot system. Section 4 describes the \(\mathcal{L}_1\) Adaptive Autopilot augmentation system. Section 5 presents simulation results, demonstrating the control system's effectiveness. The paper concludes in Section 6 with a summary of findings and their implications for AUV research.

\section{Problem Formulation}
\begin{figure}
  \centering    \includegraphics[width=.30\textwidth]{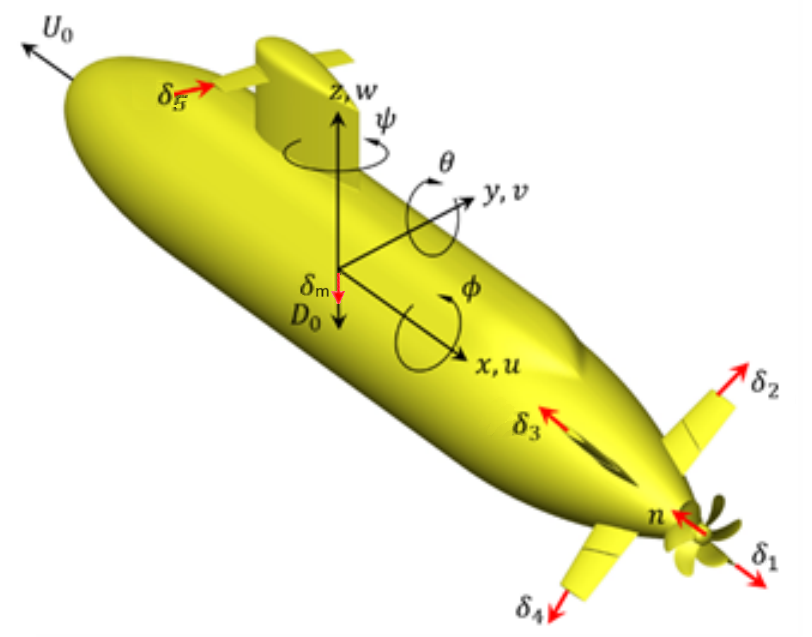}
    \caption{The axes and variables used in the BB2 model}
    \label{fig:bb2_vehicle}
\end{figure}

Consider the Joubert BB2 vehicle illustrated in Figure \ref{fig:bb2_vehicle}. The control mechanism of the vehicle primarily involves the manipulation of tail plane angles, \(\delta_1,\ldots,\delta_4\), which are arranged in an X-plane configuration, as well as the angle of the sail plane, denoted as \(\delta_5\). Similarly to \cite{overpelt2015free,rober20223d,rober2021three}, these variables can be adjusted to exert specific vertical and horizontal commands, represented by \(\delta_v\) and \(\delta_h\) respectively. The control allocation strategy is defined by the following equations:
\begin{equation}
\begin{split}
& \delta_1 = \delta_h - \delta_v, \quad
\delta_2 = -\delta_h - \delta_v, \quad
\delta_3 = -\delta_h + \delta_v, \\
& \delta_4 = \delta_h + \delta_v, \quad
\delta_5 = \delta_v.
\end{split}
\end{equation}
Additionally, the vehicle in question is equipped with a hover tank. The hover tank mass, \(\delta_m\), can be regulated to modulate the vehicle's depth dynamics. This paper primarily focuses on depth and pitch control. Consequently, our discussion centers on the application of control inputs \(\delta_v\) and \(\delta_m\) for the regulation of depth and pitch dynamics. 

Let the dynamic system of interest be governed by the following dynamic equation 
\begin{equation}
\dot{x}(t) = A x(t) + B (u(t) + f(x, t)) \label{eq:dynamics1}
\end{equation}
where \( x(t) = [z(t), \theta(t), w(t), q(t)]^\top \) represents the state vector, \(u(t) = [ \delta_v(t) , \delta_m(t) ]^\top\) is the control input vector, and \( f(x,t) \) denotes unknown nonlinear disturbances such as waves, nonlinearities, suction, etc. 
The objective is to derive control laws for \(u(t) =[ \delta_v(t) , \delta_m(t) ] \) to ensure that the parameters \( z(t) \) and \( \theta(t) \), which are governed by the dynamics presented in Equation \eqref{eq:dynamics1}, effectively follow commands denoted by \( z_{\text{cmd}}(t) \) and \( \theta_{\text{cmd}}(t) \).

Since this research focuses on the control of underwater vehicles near the surface in wave-affected environments, the design of the inputs \( \delta_v(t) \) and \( \delta_m(t) \) necessitates the cancellation of wave frequencies. This requirement is important for multiple reasons: it prevents excessive wear on actuators and aids in energy conservation. In addition, the controller should be engineered to counteract low-frequency disturbance, such as suction forces and dynamics not included in the model, to effectively eliminate errors in the steady-state performance.

\begin{figure}
\begin{tikzpicture}[block/.style={rectangle, draw, minimum height=1cm, minimum width=1cm, font=\fontsize{8pt}{10}\selectfont}, 
                    input/.style={coordinate}, 
                    output/.style={coordinate}, 
                    arrow/.style={-Stealth, font=\fontsize{8pt}{10}\selectfont}]
    \node [block] (aug) {\begin{tabular}{c} \(\mathcal{L}_1\) \\ Augmentation \end{tabular}};
    \node [block, right=0.7cm of aug] (lqr1) {LQR};
    \node [block, right=1.8cm of lqr1] (control allocation) {\begin{tabular}{c} Control \\ Allocation\end{tabular}};
    \node [block, below=1cm of control allocation] (vehicle) {Vehicle};
    \node [block, below=1cm of lqr1] (bandpass filter) {\begin{tabular}{c}Bandpass \\ Filter\end{tabular}};

    \coordinate [above left=0.3cm and 0.7cm of aug.west] (input1);
    \coordinate [below left=0.3cm and 0.7cm of aug.west] (input2);

    \draw [arrow] ([yshift=3mm]input2) -- node[above] {\( z_{\text{cmd}} \)} (aug.west); 
 \draw [arrow] ([yshift=10mm, xshift=0mm]lqr1.north) -- node[pos=0.75, right] {\( \theta_{\text{cmd}} \)} (lqr1.north); 

    \draw [arrow] ([yshift=0mm]aug.east) -- node[below] {\( z_{\text{aug}} \)} ([yshift=0mm]lqr1.west);

    \draw [arrow] ([yshift=3mm]lqr1.east) -- node[above] {\( \delta_{v} \)} ([yshift=3mm]control allocation.west);
    
    \draw [arrow] ([yshift=-3mm]lqr1.east) -- node[above] {\( \delta_{m} \)} ([yshift=-3mm]control allocation.west);
    \draw [arrow] ([yshift=-3mm]lqr1.east) -- ++(1.2cm,0) -- ++(0,-1.4cm) |- ([yshift=3mm]bandpass filter.east);
    \draw [arrow] ([yshift=3mm]lqr1.east) -- ++(1.5cm,0) -- ++(0,-2.6cm) |- ([yshift=-3mm]bandpass filter.east);

    \draw [arrow] ([xshift=3mm]bandpass filter.north) -- ++(0,0) -| node[pos=0.75, right] {\( \delta_{H_v} \)} ([xshift=3mm]lqr1.south);
   
    \draw [arrow] ([xshift=-3mm]bandpass filter.north) -- ++(0,0) -| node[pos=0.75, left] {\( \delta_{H_m} \)} ([xshift=-3mm]lqr1.south);

    \draw [arrow] ([xshift=5mm]control allocation.south) -- ++(0,0) -| node[pos=0.75, right] {\( \delta_{1..5} \)} ([xshift=5mm]vehicle.north);
    \draw [arrow] ([xshift=-5mm]control allocation.south) -- ++(0,0) -| node[pos=0.75, right] {\( \delta_{m} \)} ([xshift=-5mm]vehicle.north);

    \node [output, below=0.5cm of vehicle] (output) {};
    
    \draw [arrow] (vehicle) -- (output);

    \draw [arrow] (vehicle) -- node[above] {\( \)} (output);

\end{tikzpicture}
\caption{Controller’s Architecture } 
\label{fig:controllerarchitecture}
\end{figure}
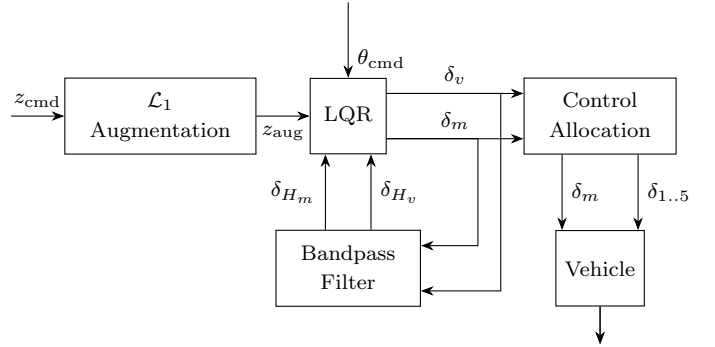
The LQR Autopilot system is specifically configured to adhere to a predetermined reference input. A critical aspect of the LQR's design is its capability to inhibit any oscillatory tendencies of the controller that correspond to the wave frequency.
The \(\mathcal{L}_1\) Adaptive Autopilot augmentation system is designed to refine the tracking reference for the autopilot system. Such a modification is crucial for ensuring that the actual AUV depth aligns precisely with their intended reference values. In other words, the \(\mathcal{L}_1\) controller modifies the reference signal in a way that guides the actual value \( z(t) \) to converge towards its respective desired target \( z_{\text{cmd}}(t) \). This results in a more accurate and efficient performance of the autopilot system. The control architecture, which consists of the LQR Autopilot and the \(\mathcal{L}_1\) Augmentation, is illustrated in Figure \ref{fig:controllerarchitecture}.

\section{LQR Autopilot}
Consider the reference signal to be tracked by the LQR autopilot defined as \( z_{\text{aug}}(t)\). Let us introduce the error state vector \( \tilde{x}(t) \), which is expressed as
\begin{equation}
\tilde{x}(t) = [z(t) - z_{\text{aug}}(t), \theta(t) - \theta_{\text{cmd}}(t), w(t), q(t)]^\top \, .
\end{equation}
The dynamics of this state vector are governed by the following differential equation:
\begin{equation}
\dot{\tilde{x}}(t) = A \tilde{x}(t) + B \left( u(t) + \tilde{f}(\tilde{x}, t) \right) \, .\label{eq:dynamics}
\end{equation}
It will become clear later that the reference \( {z}_{\text{aug}}(t) \) is dynamically adjusted to ensure convergence of the vehicle's depth to the desired command \(z_{\text{cmd}}(t)\). Nevertheless, the current focus is on the role of the LQR Autopilot in tracking this augmented command signal.

A preliminary strategy might involve examining a simplified version of the system, i.e., 
\begin{equation}
\dot{\tilde{x}}(t) = A \tilde{x}(t) + B u(t) \, ,\label{eq:dynamicslinear}
\end{equation}
where external disturbances and unmodelled dynamics are ignored. Subsequently, one could formulate an LQR state-feedback control law as follows:
\begin{equation}
u(t) = -K_{\text{LQR}} \tilde{x}(t) , \label{eq:naiveLQR}
\end{equation}
aiming to minimize a quadratic cost function
\begin{equation}
J = \int_0^\infty \left( \tilde{x}(t)^\top Q \tilde{x}(t) + u^\top (t) R u(t) \right) dt \, .
\end{equation} 
However, this approach has limitations. While it can stabilize the closed-loop system under certain disturbance assumptions, it inherently includes an \emph{internal model} of the disturbance within the control system. To elaborate, if \( \tilde{f}(\cdot) \) contains wave dynamics, the controller \( u(t) \) integrates a model of these waves (see the Internal Model Principle in \cite{francis1976internal}). These induced oscillations in the control input present several drawbacks. They can potentially result in actuator damage, elevate noise levels, and escalate energy consumption. Thus, it is essential to develop a strategy that effectively mitigates or eliminates these oscillations from the control input to enhance system performance and reliability. 

To attenuate these undesired dynamics, the application of a filter is a common strategy. Specifically, the controller defined in Equation \eqref{eq:naiveLQR} is modified as
\begin{equation} 
u_F(s) = F(s)u(s)\label{eq:filteredLQR}
\end{equation} 
where \(u(s)\) is the Laplace transform of \eqref{eq:naiveLQR} and \(F(s)\) is a filtering function, such as a notch/band-stop filter, designed to attenuate specific frequency components of the input. However, this approach has a notable limitation: the introduction of the filter to the open-loop system introduces a negative phase margin, thereby adversely affecting the robustness of the closed-loop system.

Given the assumption of known or estimated wave frequencies, we adopt the following approach to attenuating the disturbance. 
The system is represented through an augmented state space model, incorporating additional states representing the inputs amplified at the pertinent frequencies. An LQR controller is then devised, specifically tailored to minimize the input at these frequencies. This minimization is achieved by employing a band-pass filter, which selectively amplifies the input signal at the targeted frequencies, thus enhancing the system's sensitivity to wave-induced dynamics. As illustrated in Figure \ref{fig:controllerarchitecture}, this methodology does not incorporate a filter in the open-loop system, thereby reducing the potential impact of wave frequency cancellation on the closed-loop system's robustness. Further details on this approach are provided in the remainder of this section.

The amplified inputs are represented by the following:
\begin{equation}
\delta_{H_v}(s) = H_v(s) \delta_v(s) \, , \quad \delta_{H_m}(s) = H_m(s) \delta_m(s) 
\end{equation}
where \(H_v(s)\) and \(H_m(s)\) represent band-pass filters, the design of which should be tailored based on factors such as wave frequency and sea state conditions.
These inputs can be expressed in state space as follows:
\begin{equation}
\left\{
\begin{aligned} \label{eq:ss.filter_v}
\dot{x}_{H_v}(t) &= A_{H_v} x_{H_v}(t) + B_{H_v} \delta_v(t)  \\
\delta_{H_v}(t) &= C_{H_v} x_{H_v}(t) + D_{H_v} \delta_v(t)
\end{aligned}
\right.
\end{equation}
\begin{equation}
\left\{
\begin{aligned} \label{eq:ss.filter_m} 
\dot{x}_{H_m}(t) &= A_{H_m} x_{H_m}(t) + B_{H_m} \delta_m(t) \\
\delta_{H_m}(t) &= C_{H_m} x_{H_m}(t) + D_{H_m} \delta_m(t) 
\end{aligned}
\right.
\end{equation}

Therefore, the systems under consideration can be described as:
\begin{equation}
\left\{
\begin{aligned}
\dot{\tilde{x}}(t) &= A \tilde{x}(t) + Bu(t)  \\
\dot{{x}}_{H_v}(t) &= A_{H_v} {x}_{H_v}(t) + B_{H_v} \delta_v(t)  \\
\dot{{x}}_{H_m}(t) &= A_{H_m} {x}_{H_m}(t) + B_{H_m} \delta_m(t)  \\
\end{aligned}
\right.
\end{equation}
which can be restructured in matrix form as:
\begin{equation} \label{eq:augfilersys}
\dot{\hat{x}}(t) = \hat{A} \hat{x}(t) + \hat{B} u(t) \end{equation}
where \(\hat{x}(t) = [\tilde{x}(t)^\top, {x}_{H_v}(t)^\top, {x}_{H_m}(t)^\top]^\top\),
\begin{equation*}
\hat{A} = \begin{bmatrix}
A & 0 & 0 \\
0 & A_{H_v} & 0 \\
0 & 0 & A_{H_m}
\end{bmatrix}, \quad
\hat{B} = \begin{bmatrix}
B \\
[B_{H_v} \; 0] \\
[0 \; B_{H_m}]
\end{bmatrix} 
\end{equation*}

An LQR controller is then designed to minimize the cost function:
\begin{equation}
\begin{aligned}
J = & \int_0^\infty \left( \tilde{x}(t)^\top Q_1 \tilde{x}(t) + \delta_{H_v}(t)^\top Q_{2,v} \delta_{H_v}(t) \right. \\
&  + \left. \delta_{H_m}(t)^\top Q_{2,m} \delta_{H_m}(t) + \begin{bmatrix} \delta_v(t), \delta_m(t) \end{bmatrix} R \begin{bmatrix} 
\delta_v(t) \\ \delta_m(t) \end{bmatrix} \right)dt
\end{aligned}
\end{equation}
Utilizing Equations \eqref{eq:ss.filter_v} and \eqref{eq:ss.filter_m}, the cost function can be expanded as follows:
\begin{equation}
\begin{aligned}
J  & =  \int_0^\infty \left( \tilde{x}(t)^\top Q_1 \tilde{x}(t) \right. \\
& \quad + (x_{H_v}(t)^\top C_{H_v}^\top + \delta_v(t) D_{H_v}^\top) Q_{2,v} (C_{H_v} x_{H_v}(t) \\ 
& \quad + D_{H_v} \delta_v(t)) + \begin{bmatrix} \delta_v(t), \delta_m(t) \end{bmatrix} R \begin{bmatrix} \delta_v(t) \\ \delta_m(t) \end{bmatrix} \\
& \quad + (x_{H_m}(t)^\top C_{H_m}^\top + \delta_m(t) D_{H_m}^\top) Q_{2,m} (C_{H_m} x_{H_m}(t) \\ 
& \left. \quad + D_{H_m} \delta_m(t)) \right)dt
\end{aligned}
\end{equation}

\begin{equation}
\begin{aligned}
J  & = \int_0^\infty \left( \hat{x}(t)^\top
\begin{bmatrix} Q_1 & 0 & 0 \\ 0 & C_F^\top Q_{2,v} C_F & 0 \\ 0 & 0 & C_F^\top Q_{2,m} C_F \end{bmatrix}
\hat{x}(t) \right. \\
& + 2 \hat{x}(t)^\top
\begin{bmatrix} 0 & 0 \\ C_F^\top Q_{2,v} D_F & 0 \\ 0 & C_F^\top Q_{2,m} D_F \end{bmatrix}
u(t) \\
& + \left. u(t)^\top \left( R +
\begin{bmatrix} D_F^\top Q_{2,v} D_F & 0 \\ 0 & D_F^\top Q_{2,m} D_F \end{bmatrix} \right)
u(t) \right)dt
\end{aligned}
\end{equation}
with dynamics governed by \eqref{eq:augfilersys}. This cost and state space dynamics lead to the LQR control law
\begin{equation} \label{eq:LQRcontroller}
u(t) = -\hat{K}_{\text{LQR}} \hat{x}(t)
\end{equation} 
which is aimed at attenuating wave frequencies.

\section{\(\mathcal{L}_1\) Augmentation}
Given the depth command \(z_{\text{cmd}}(t)\) as reference input, the subsequent implementation of the \(\mathcal{L}_1\) augmentation ensures this command is accurately executed. 
With the LQR controller implemented, the depth tracking error of the autopilot, \(z(t)-z_{\text{aug}}(t)\), is assumed to be stable and bounded. The dynamics of the closed-loop system, comprising the vehicle and the autopilot, are mathematically represented as:
\begin{equation} \label{eq:TF_LQRautopilot}
\mathcal{G}_p: \,
z(s)  = G_z(s) (z_\text{aug}(s) + d_z(s)) . 
\end{equation}

where \(G_z(s)\) is unknown strictly proper and stable function, and \(d_z(s)\) is the Laplace transform of time-varying uncertainties and disturbance signals. With a slight abuse of notation, the above system can be written in state space as follows
\begin{equation} \label{eq:APsys_z_}
       \begin{cases}
        \dot{{x}}_z(t) = {A_{z}} {x}_z(t) + {B_{z}} ({z_{\text{aug}}}(t)+{d}_z({x}_z(t),t)) \\
        z(t) = {C_{z} x_z}(t),  \\
    \end{cases}
\end{equation}

where the tuple $\left\{{A_{z}},{B_{z}},{C_{z}}\right\}$  is the minimum realization of \({G}_z\). The core idea is to dynamically adjust the variable \(z_{\text{aug}}(t)\), thereby ensuring the convergence of \(z(t)\) towards the prescribed reference value \(z_{\text{cmd}}(t)\). This method of adaptive autopilot augmentation bears similarity to approaches documented in \cite{kaminer2010path, rober20223d, MACLIN2024105792} The principle mirrors practical scenarios in manned vehicle navigation, where pilots employ control interfaces, such as joysticks, to maintain desired depth. The feedback loop involves continuous monitoring of and adaptation to the vehicle's response to achieve the targeted depth, paralleling the adaptive control mechanisms proposed here.

We now introduce the desired system dynamics as follows:
\begin{equation} \label{eq:TFdesired}
z_m(s)  = M_z(s) z_\text{cmd}(s),
\end{equation}
Here, \(M_z(s)\) represents the desired transfer function to be devised by the control designer to fulfill specific performance criteria, with the condition that \(M_z(0) =  1\). Furthermore, \(z_\text{cmd}(s)\) denotes the Laplace transform of the corresponding command signal. 

We observe that Equation \eqref{eq:TF_LQRautopilot} can be reformulated as:
\begin{equation} \label{eq:TFactualz}
z(s) = M_z(s) (z_\text{aug}(s) + \sigma_z(s)),
\end{equation}
where the uncertainties encompassing \(G_z(s)\) and \(d_z(s)\) are encapsulated within the term \(\sigma_z(s)\). I.e.,
\begin{equation}
\sigma_z(s) = \frac{(G_z(s)-M_z(s))z_\text{aug}(s) + G_z(s)d_z(s)}{M_z(s)}.
\end{equation}
The underlying strategy involves adjusting \(z_\text{aug}(s)\) such that the system in Equation \eqref{eq:TFactualz} emulates the behavior of Equation \eqref{eq:TFdesired}. To this end, the controller estimates the uncertainty \(\sigma_z(s)\) and modifies the control input \(z_\text{aug}(s)\) to counteract these uncertainties.

The state-space representation of system \eqref{eq:TFactualz} is given by:
\begin{equation} \label{eq:APsys_z}
    \begin{cases}
        \dot{{x}}_z(t) = {A_{m}} {x}_z(t) + {B_{m}} ({z_{\text{aug}}}(t) + {\sigma}_z(x_z,t)), \\
        z(t) = {C_{m} x_z}(t),
    \end{cases}
\end{equation}
where \(\left\{{A_{m}},{B_{m}},{C_{m}}\right\}\) constitutes the minimal realization of \({M}_z(s)\). Consequently, \(A_m\) is Hurwitz, the tuple \(\left\{{A_{m}},{B_{m}},{C_{m}}\right\}\) is both controllable and observable, and it holds that \( -C_m A_m^{-1}B_m = 1\).

Before we proceed with the design of the controller, the following assumptions are made. 
\begin{asm} \label{asm:r}
	The reference command is bounded as follows: $$||z_\text{cmd}(t)||_\infty \leq M_r \, , \quad \forall t \geq 0 .$$
\end{asm}
\begin{asm} \label{asm:f} 
	For any $\delta>0$ there exists $F_\delta$ and $L_0$ such that
	\begin{equation*}
	\begin{split}   
	\Vert \sigma_z(x_2,t)-\sigma_z(x_1,t) \Vert_\infty  & \leq F_{\delta} \Vert x_2 - x_1 \Vert_\infty , \\
	\Vert \sigma_z(0,t) \Vert_\infty & \leq L_0    
	\end{split} 
	\end{equation*}  
	hold uniformly for all $\Vert x_i \Vert \leq \delta $, $i \in \{ 1 , 2 \}$, $t \geq 0$.
\end{asm}

\subsection{Control law} \label{sec:controller}
In this section, we present the sampled-data controller based on the control architecture introduced in \cite{jafarnejadsani2019l1}. The controller comprises three principal components: an output predictor, an adaptation law, and a control law, detailed as follows.

\textbf{Output Predictor.} The discrete-time output predictor is represented by:
{ \footnotesize
\begin{equation} \label{eq:predictor}
\begin{split} 
\hat{x}_z[i+1] & = e^{A_mT_s}\hat{x}_z[i] + A_m^{-1}(e^{A_mT_s}-I)(B_m z_{\text{aug}}[i] + \hat{\sigma}_z[i]), \\
\hat{z}[i] & = C_m \hat{x}_z[i], \quad \hat{x}_z[0] = C_m^{\dagger} z_0,
\end{split}
\end{equation}
}
where \(T_s\) is the sampling rate, \(\hat{x}_z[i] \in \mathbb{R}^3\) is the state, \(\hat{z}[i] \in \mathbb{R}\) is the output, \(z_0\) is the initial output, and \(\hat{\sigma}_z[i]\) and \(z_\text{aug}[i] \in \mathbb{R}\) are the estimated uncertainty and control input, respectively. The predictor replicates the closed-loop dynamics, substituting the unknown function \(\sigma_z(x_z,t)\) with the estimate \(\hat{\sigma}_z[i]\).

\textbf{Adaptation Law.} The estimate \(\hat{\sigma}_z[i]\) is computed as:
\begin{equation}
\hat{\sigma}_z [i] = -\Phi^{-1}(T_s) e^{\Lambda A_m \Lambda^{-1} T_s} \begin{bmatrix} 1 \\ 0 \\ 0 \end{bmatrix} (\hat{z}[i]-z[i]),
\end{equation}
where \(z[i] = z(iT_s)\), \(i \in \mathbb{Z}^{+}\), and \(\hat{z}[i]\) is as in Equation \eqref{eq:predictor}. The matrix \(\Lambda\) is defined by:
\begin{equation}
\Lambda = \begin{bmatrix} C_m \\ D \sqrt{P} \end{bmatrix},
\end{equation}
with \(P = \sqrt{P}^\top \sqrt{P}\) and \(D\) satisfying \(D(C_m (\sqrt{P})^{-1})^\top=0\). The matrix \(\Phi(T_s)\) is given by:
\begin{equation}
\Phi(T_s) = \int_0^{T_s} e^{\Lambda A_m \Lambda^{-1} (T_s-\tau)} \Lambda d\tau.
\end{equation}

\textbf{Control Law.} The control input \(z_{\text{aug}}(t)\) for Equation \eqref{eq:APsys_z} is defined over discrete intervals:
\begin{equation}
z_{\text{aug}}(t) = z_{\text{aug}}[i], \quad t \in [iT_s, (i+1)T_s), \quad i \in \mathbb{Z}^{+},
\end{equation}
where \(z_{\text{aug}}[i]\) is determined by:
\begin{equation} \label{eq:controlinput}
\begin{split} 
x_u[i+1] & = e^{A_oT_s}x_u[i] + A_o^{-1}(e^{A_oT_s}-I)(B_o e^{-A_m T_s} \hat{\sigma}_z[i]), \\
z_\text{aug}[i] & = z_\text{cmd}[i] - C_o x_u[i], \quad x_u[0] = 0,
\end{split}
\end{equation}
with \(z_{\text{cmd}}[i] = z_{\text{cmd}}(iT_s)\) and \(i \in \mathbb{Z}^+\). The tuple \((A_o,B_o,C_o)\) is the minimal state-space realization of the transfer function:
\begin{equation} \label{eq:OS}
O(s) = C(s)M_z^{-1}(s)C_m(sI - A_m)^{-1},
\end{equation}
where \(C(s)\) is a strictly proper stable transfer function satisfying \(C(0) = 1\).

Finally, the controller at hand consists of Equations \eqref{eq:predictor}-\eqref{eq:OS} and is subject to the following two conditions:
\begin{equation} \label{eq:condproper}
C(s)M_z^{-1}(s) \quad \text{is proper,} 
\end{equation}
and, for a given $\rho_0$, there exists $\rho_r$ such that the following $\mathcal{L}_1$ norm condition holds
\begin{equation}  \label{eq:L1condition}
\Vert G(s) \Vert_{\mathcal{L}_1} < \frac{\rho_r - \rho_1 - \rho_2}{L_{\rho_r}\rho_r + L_0} , 
\end{equation}
where 
\begin{equation} \label{eq:TFs}
\begin{split}  
& G(s) = H_0(s) (I - C(s)) , \quad
H_0(s)  = (sI-A_m)^{-1}B_m \, , \\
& \rho_1  = \Vert s(sI-A_m)^{-1}-sH_1(s)H_{\text{in}}\Vert_{\mathcal{L}_1} \rho_0 , \\
& \rho_2 = \Vert H_0(s) \Vert_{\mathcal{L}_1} M_r , 	\\    
& H_{\text{in}}(s)  = C_m(sI-A_m)(I-C_m^\dagger C_m) , \\
& H_1(s)  = H_0(s)C(s)M_z^{-1}(s) \, , \\
& L_{\rho_r}  = \frac{\bar{\gamma}_0 + \rho_r}{\rho_r}(F_{\bar{\gamma}_0 + \rho_r} + \Vert K \Vert_\infty) ,
\end{split}
\end{equation} 
$\bar{\gamma}_0$ is an arbitrarily small positive constant, $F_{\delta}$, $L_0$ and $M_r$ are introduced in Assumptions \ref{asm:r} and \ref{asm:f}, and \(K\) is selected such that \(A_z - B_zK \) is Hurwitz. 


\subsection{Controller performance bounds} \label{sec:perfbounds}
In order to analyze the performance of the controller presented in the previous section, we introduce the following reference system
\begin{equation} \label{eq:referencesystem}
\begin{split} 
\dot{x}_{\text{ref}}(t) & = A_z{x}_{\text{ref}}(t)+ B_z ({u}_{\text{ref}}(t)+\sigma_z(t,{x}_{\text{ref}}(t))), \\
{u}_{\text{ref}}(s) & = z_{\text{cmd}}(s) - C(s) {\sigma}_{\text{ref}}(s), \\
{z}_{\text{ref}}(t) & = C_z {x}_{\text{ref}}(t), \quad {x}_{\text{ref}}(0) = {x}_{0} ,
\end{split} 
\end{equation}
where 
$$
\sigma_{\text{ref}} (s) = \omega_{\text{ref}}(s) + H_{\text{in}}(s)x_0 ,
$$
$\omega_{\text{ref}}(s)$ is the Laplace transform of 
$$
\omega_{\text{ref}}(t) = Kx_{\text{ref}}(t) +\sigma_z(x_{\text{ref}}(t),t) . 
$$

Notice that the output of the reference system can be written as
\begin{equation} \label{eq:refoutput} 
\begin{split}
z_{\text{ref}}(s) & = M_z(s) z_{\text{cmd}}(s) + M_z(s)(I-C(s))\sigma_{\text{ref}}(s) \\ 
& \qquad + C_m(sI-A_m)^{-1}C_m^\dagger y_0 ,
\end{split} 
\end{equation} 
i.e. the reference system in Equation \eqref{eq:referencesystem} mitigates only the uncertainty $\sigma_{\text{ref}}(t)$ that is within the bandwidth of $C(s)$. Moreover, since $C(0)=1$, application of the Final Value Theorem implies that the reference system recovers the desired response introduced in Equation \eqref{eq:TFdesired}.

\begin{rem}
Notice that the control input of the reference system depends on the uncertainty $\sigma_z(x_{\text{ref}}(t),t)$ and the unknown state vector $x_{\text{ref}}(t)$, and thus is not implementable. In turn, the reference system in Equation \eqref{eq:referencesystem} is introduced only for the purpose of performance analysis. 
\end{rem}

The following lemma establishes stability results and performance bounds concerning the reference system. 

\begin{lem} \label{lem:refstable}
Consider the closed-loop reference system in Equation \eqref{eq:referencesystem} subject to \eqref{eq:condproper} and \eqref{eq:L1condition}. If $\Vert x_0 \Vert_{\infty} \leq \rho_0$, then
\begin{equation}
\Vert x_{\text{ref}}(t) \Vert_{\mathcal{L}_\infty} \leq \rho_{\text{r}} \, \quad 
\Vert u_{\text{ref}}(t) \Vert_{\mathcal{L}_\infty} \leq \rho_{\text{ur}}
\end{equation}
where $\rho_{\text{r}}$ was introduced in Equation \eqref{eq:L1condition} and 
\begin{equation*} 
\begin{split} 
\rho_{\text{ur}} & = \Vert C(s) \Vert_{\mathcal{L}_1}(L_{\rho_{\text{r}}} \rho_{\text{r}}+L_0) + \Vert sC(s)M_z^{-1}(s)H_{\text{in}}(s) \Vert_{\mathcal{L}_1} \rho_0 \\
	& \qquad + \Vert M_r \Vert_{\infty}
\end{split}
\end{equation*} 
\end{lem}

Finally, in the following theorem we present the stability and performance bounds of the closed-loop autopilot augmentation system. 

\begin{thm} \label{thm:mainresult}
	Consider the system given by Equation \eqref{eq:APsys_z} and the control laws proposed in Equations \eqref{eq:predictor}-\eqref{eq:OS} subject to conditions \eqref{eq:condproper} and \eqref{eq:L1condition}. If $\Vert x_0 \Vert_{\infty} \leq \rho_0$, then 
	\begin{equation}
	\Vert x_{\text{ref}}(t)-x_z(t) \Vert_{\mathcal{L}_\infty} \leq \gamma_{x} ,
	\end{equation}
	and
	\begin{equation}
	\Vert u_{\text{ref}}(t) - z_{\text{aug}}(t) \Vert_{\mathcal{L}_\infty} \leq \gamma_{u} ,
	\end{equation}
	with 
	$$
	\lim_{T_s \to 0} \gamma_x = \lim_{T_s \to 0} \gamma_u = 0 \, .
	$$
\end{thm}

\textbf{Proof.} Due to constraints in manuscript length, the detailed proofs of Lemma \ref{lem:refstable}  and Theorem \ref{thm:mainresult} are omitted in this document. Nevertheless, the proofs align with the ones outlined in \cite{jafarnejadsani2019l1} and \cite{rober20223d}.

Theorem \ref{thm:mainresult} indicates that the system described in Equation \eqref{eq:APsys_z} can be aligned closely with the reference system outlined in Equation \eqref{eq:referencesystem} by lowering the sampling rate \(T_s\). Additionally, as indicated by Equation \eqref{eq:refoutput}, the variable output \(z_{\text{ref}}(t)\) follows the desired output \(z_m(t)\) (Equation \eqref{eq:TFdesired}). Therefore, based on Theorem \ref{thm:mainresult}, we conclude that the output \(z(t)\) tracks the desired output \(z_m(t)\) in both transient and steady-state. The performance bounds can be reduced by appropriately selecting \(T_s\) and \(C(s)\).

In the control architecture presented, the term $C(s)$ denotes the low-pass filter applied to the control input. This filter plays a crucial role in attenuating high-frequency components present in the adaptive control signals. The selection of this filter must adhere to the criteria outlined in conditions \eqref{eq:condproper} and \eqref{eq:L1condition}. The incorporation of this filter enables the adaptation to address disturbances occurring at lower frequencies, commonly encompassing phenomena such as suction forces and unmodeled dynamics. For more discussion on filter design strategies that balance robustness and performance, readers are referred to \cite{jafarnejadsani2017optimized,hovakimyan2010L1}.

The inverse of the sampling time $T_s$ plays the role of adaptation gains found in typical continuous-time adaptive control architectures. Similarly to the $\mathcal{L}_1$ piecewise-constant adaptive laws introduced in \cite{hovakimyan2010L1}, this parameter can be directly related with the sampling rate of the CPU. 
The sampling time should always be selected as low as possible, within the limits of the CPU, in order to achieve high controller performance.

\section{Numerical results}
The analysis presented in this section focuses on depth keeping and changing maneuvers in the presence of monochromatic waves, specifically in sea state 5. These maneuvers are conducted under two velocity regimes: at a lower velocity of 2 m/s (denoted as Scenario 1) and at a higher velocity of 5 m/s (referred to as Scenario 2). Within each scenario, the study evaluates three distinct controller architectures. The first configuration is characterized by the absence of augmentation, i.e., \(z_{\text{aug}}(t) = z_{\text{cmd}}(t)\), and the lack of a filtering mechanism (Case 1). In this setup, the controller as outlined in Equation \eqref{eq:naiveLQR} is employed with 

\begin{equation}
\scalebox{0.9}{
$Q = \begin{bmatrix}
50 & 0 & 0 & 0 \\
0 & 50 & 0 & 0 \\
0 & 0 & 2000 & 0 \\
0 & 0 & 0 & 2000
\end{bmatrix},
\quad 
R = \begin{bmatrix}
500 & 0 \\
0 & 0.1
\end{bmatrix}$
}
\end{equation}
The second configuration (Case 2) involves a setup without augmentation but includes a filtering process, implementing the controller specified in Equation \eqref{eq:LQRcontroller} with 
\begin{equation}
\scalebox{0.9}{
$Q_1 = \begin{bmatrix}
50 & 0 & 0 & 0 \\
0 & 50 & 0 & 0 \\
0 & 0 & 2000 & 0 \\
0 & 0 & 0 & 2000
\end{bmatrix}$
}
\end{equation}
\begin{equation}
\scalebox{1.0}{
$Q_{2,v} = 1 \quad Q_{2,m} = 10^{-5} \quad R = \begin{bmatrix}
500 & 0 \\
0 & 0.1
\end{bmatrix}$
}
\end{equation}
Additionally, the bandpass filters $H_{v}(s)$ and $H_{m}(s)$ are implemented as the inverse of \(8\)th-order notch filters, specifically tuned to the frequencies of the waves. 
The final configuration (Case 3) integrates both augmentation and filtering into the control architecture. The \(\mathcal{L}_1\) augmentation parameters are selected as
\begin{equation}
\scalebox{1.0}{
$M_z(s) = \frac{0.0064}{s^2 + 0.16s + 0.0064}$}
\end{equation}
\begin{equation}
\scalebox{1.0}{
$C(s) = \frac{0.01^3}{(s + 0.01)^3} \, \quad
T_s = 0.05 $}
\end{equation}

\subsection{Scenario 1 - Low Speed}
The dynamic model for depth and pitch of the Joubert BB2 vessel, navigating at a speed of \(2\) m/s, is represented by Equation \eqref{eq:dynamics1} with 
\begin{equation}
\scalebox{0.85}{
$A = \begin{bmatrix}
9.056 \cdot 10^{-14} & 1.999 & 1 & -9.466 \cdot 10^{-12} \\
-1.527 \cdot 10^{-13} & 1.59 \cdot 10^{-14} & -2.135 \cdot 10^{-11} & 1 \\
0.00013 & -3.04 \cdot 10^{-7} & -0.036 & -1.144 \\
7.005 \cdot 10^{-7} & -0.0149 & -0.00258 & -0.095
\end{bmatrix} $}
\end{equation}
\begin{equation}
\scalebox{0.85}{
$B = \begin{bmatrix}
3.257 \cdot 10^{-14} & -3.078 \cdot 10^{-19} \\
-9.036 \cdot 10^{-14} & -1.156 \cdot 10^{-18} \\
-0.00015 & -2.219 \cdot 10^{-6} \\
6.444 \cdot 10^{-5} & 4.595 \cdot 10^{-17}
\end{bmatrix} $}
\end{equation}
The matrices are derived from the Reduced Order Model (ROM) outlined in \cite{martin2022reduced}, employing the Matlab Simulink linearization toolbox for computation.

The submarine starts its operation at a depth of 15 meters. Given the considerable size of the vehicle, this places it very close to the water’s surface. At this shallow depth, the AUV is significantly influenced by disturbances from near-surface dynamics where it maintains this depth for a predetermined duration.  It then shifts to 20 meters, stabilizing there for a period to assess controller stability, before descending to 50 meters, where it remains for a while in the absence of surface suction effects. Subsequently, it returns to 20 meters, sustains that depth, and finally ascends back to 15 meters. This pattern, alternating and holding depths at 15, 20, and 50 meters, highlights the varying influence of surface suction on AUV operations.

\begin{figure}
  \centering    \includegraphics[width=.48\textwidth]{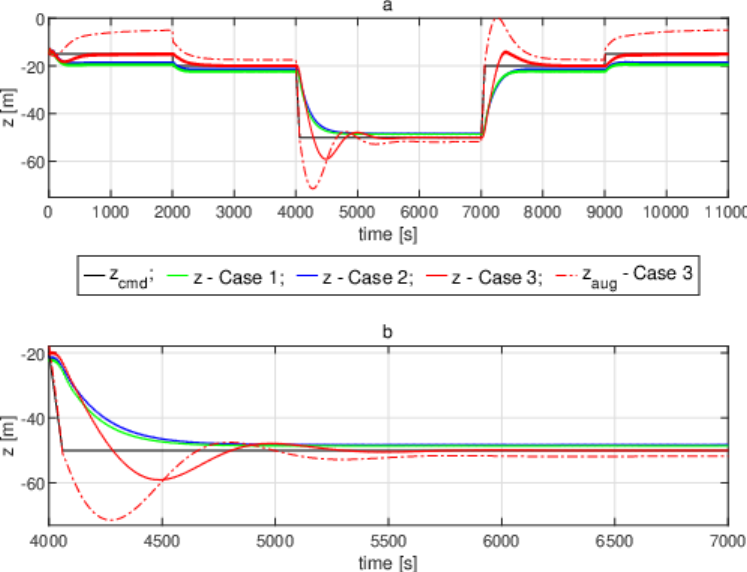}
    \caption{Scenario 1, Cases 1,2,3. Depth change. 
    a - depth change for whole scenario;
    b - depth change for the specified location.}
    \label{fig:z_v2}
\end{figure}

\begin{figure}
  \centering    \includegraphics[width=.48\textwidth]{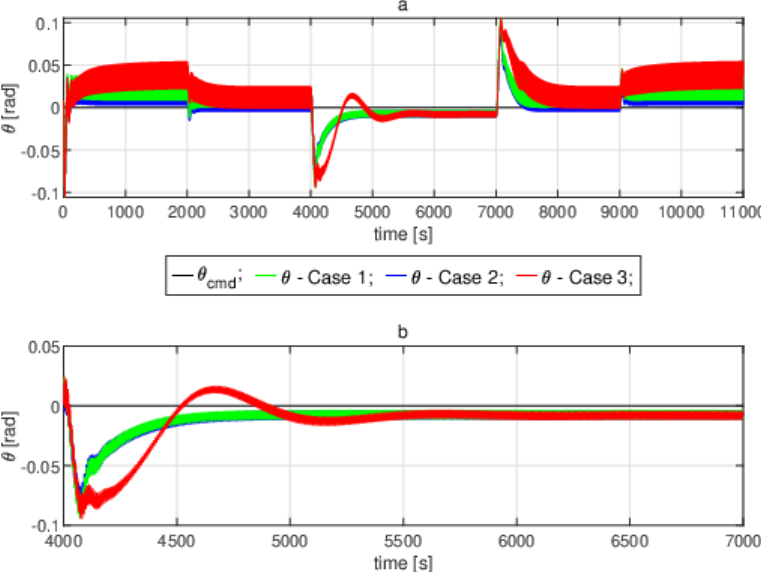}
    \caption{Scenario 1, Cases 1,2,3. $\theta$ change. a - $\theta$ change for whole scenario;
    b - $\theta$ change for the specified location.}
    \label{fig:theta_v2}
\end{figure}

\begin{figure}
  \centering    \includegraphics[width=.48\textwidth]{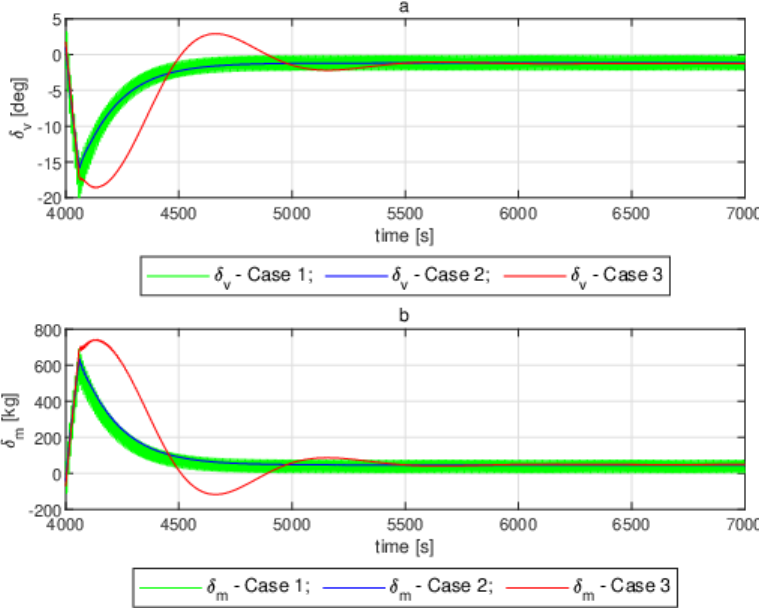}
    \caption{Scenario 1, Cases 1,2,3. $\delta_{v}$, $\delta_{m}$ change for the specified location. a - $\delta_{v}$; 
    b - $\delta_{m}$.}
    \label{fig:dv_dm_v2}
\end{figure}

Figures \ref{fig:z_v2}, \ref{fig:theta_v2} and \ref{fig:dv_dm_v2} present a series of six interconnected charts. Figure \ref{fig:z_v2}a illustrates the depth-changing maneuver for all three cases, depicting the AUV's sequential depth transitions. Figure \ref{fig:z_v2}b details the specific region traversed by the AUV during the simulation timeframe between 4000 and 7000 seconds. Figures \ref{fig:theta_v2}a and \ref{fig:theta_v2}b show the change in $\theta$ during the whole simulation and the specific timeframe respectively. Figures \ref{fig:dv_dm_v2}a and \ref{fig:dv_dm_v2}b, respectively, highlight the effects on the planes ($\delta_{v}$) and the hover tank ($\delta_{m}$).

In Case 1, pronounced fluctuations are evident in both the plane orientations and the hover tank, attributable to the disturbances caused by sea state 5 conditions. Additionally, the AUV faces challenges in reaching the specified depths during its depth-changing maneuvers.
For Case 2, incorporating bandpass filters into the plane systems and hover tank results in a reduction of the oscillatory effects induced by sea disturbances on the operational actuators and hover tank. Nevertheless, the system still encounters some difficulties in navigating to the desired depths.
In Case 3, the application of bandpass filters in tandem with adaptive augmentation method improves performance. Owing to the adjustment made by $z_{aug}(t)$, actual depth $z(t)$ achieves and maintains the intended depth \(z_{\text{cmd}}\).

\subsection{Scenario 2 - High Speed}

Scenario 2 mirrors Scenario 1, with the only difference being the increased speed of the AUV. Whereas in the previous instance the speed was 2 m/s, it is now increased to 5 m/s.
\begin{figure}
  \centering    \includegraphics[width=.48\textwidth]{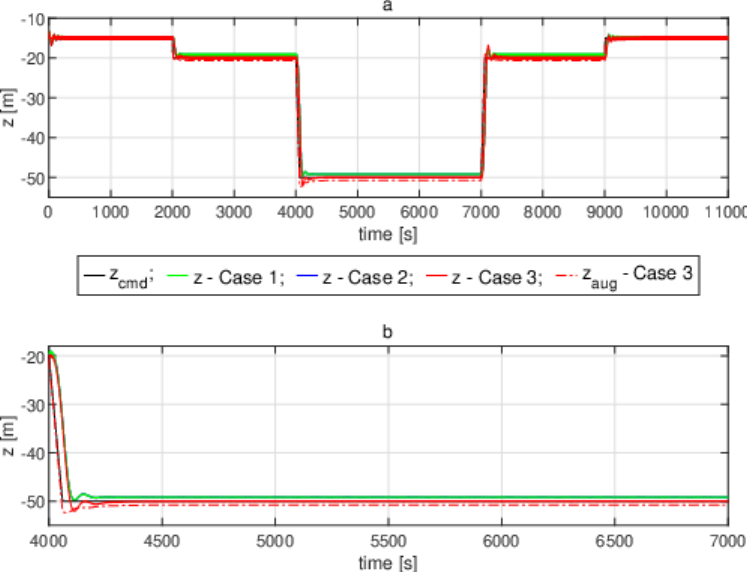}
    \caption{Scenario 2, Cases 1,2,3. Depth change. a - depth change for whole scenario; b - depth change for the specified location.}
    \label{fig:z_v5}
\end{figure}
 \begin{figure}
  \centering    \includegraphics[width=.48\textwidth]{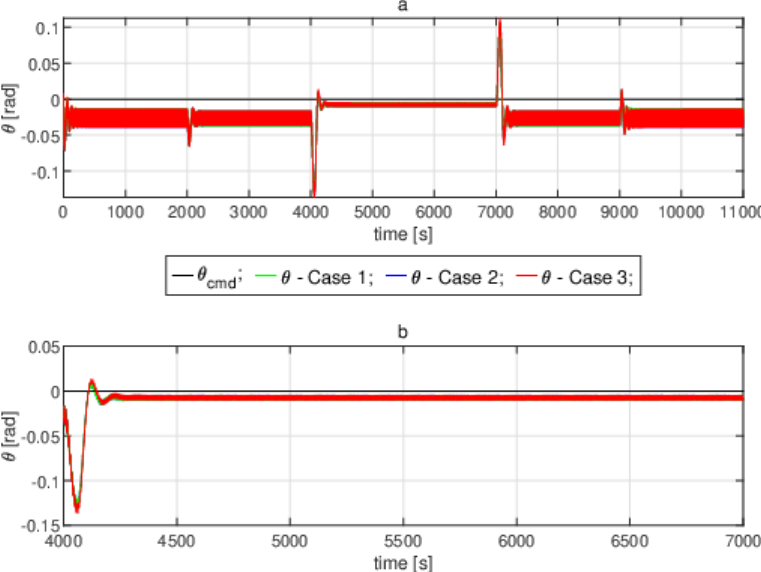}
    \caption{Scenario 2, Cases 1,2,3. $\theta$ change. a - $\theta$ change for whole scenario; b - $\theta$ change for the specified location.}
    \label{fig:theta_v5}
\end{figure}
\begin{figure} 
  \centering    \includegraphics[width=.48\textwidth]{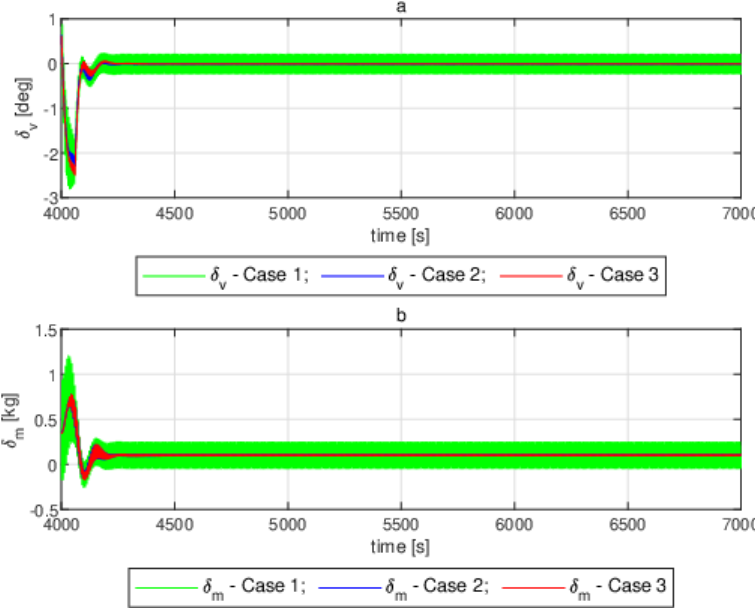}
    \caption{Scenario 2, Cases 1,2,3. $\delta_{v}$, $\delta_{m}$ change for the specified location. a - $\delta_{v}$; b - $\delta_{m}$.}
    \label{fig:dv_dm_v5}
\end{figure}
The dynamic model, describing the depth and pitch behaviors of the Joubert BB2 vessel while navigating at a velocity of 5 m/s is characterized by the following state space matrices (also obtained via Matlab Simulink)

\begin{equation}
\scalebox{0.85}{
$A = \begin{bmatrix}
-9.613 \cdot 10^{-13} & 4.999 & 1 & 1.122 \cdot 10^{-10} \\
1.452 \cdot 10^{-12} & -5.139 \cdot 10^{-14} & -1.178 \cdot 10^{-10} & 1 \\
0.00084 & -3.04 \cdot 10^{-7} & -0.1016 & -2.7 \\
4.379 \cdot 10^{-6} & -0.0149 & -0.00572 & -0.244
\end{bmatrix} $}
\end{equation}
\begin{equation}
\scalebox{0.85}{
$B = \begin{bmatrix}
1.06 \cdot 10^{-12} & -1.104 \cdot 10^{-17} \\
-4.927 \cdot 10^{-12} & -6.124 \cdot 10^{-18} \\
-0.000969 & -2.219 \cdot 10^{-6} \\
0.0004 & 2.403 \cdot 10^{-16}
\end{bmatrix}.$}
\end{equation}




Figures \ref{fig:z_v5}a and \ref{fig:z_v5}b illustrate the test maneuver.
The Figure \ref{fig:z_v5}b highlights the simulation data captured between 4000 and 7000 seconds. Correspondingly, Figures \ref{fig:theta_v5}a and \ref{fig:theta_v5}b present data within the same timeframe, but they concentrate on changes in the pitch angle \(\theta\). The pair of Figures \ref{fig:dv_dm_v5}a and \ref{fig:dv_dm_v5}b demonstrate the effects exerted by the fin control \(\delta_{v}\) and the hover tank \(\delta_{m}\), respectively.

\begin{figure} 
  \centering    \includegraphics[width=.48\textwidth]{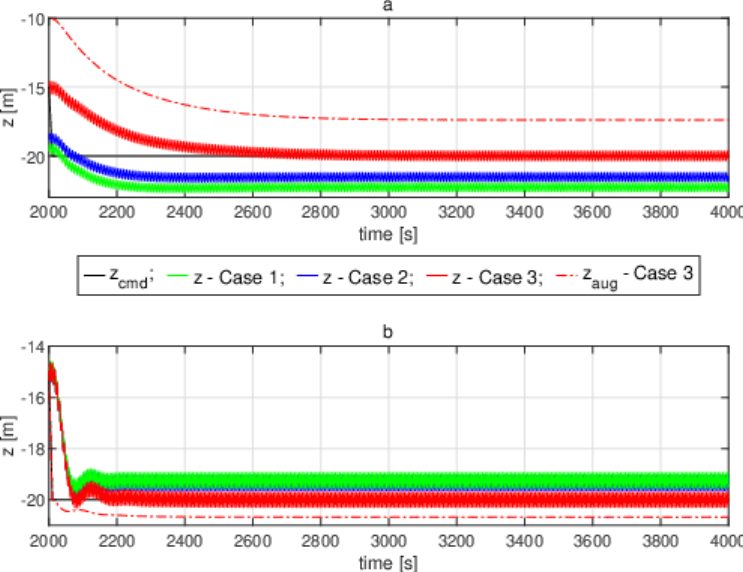}
    \caption{Comparison between Scenario 1 and Scenario 2 , Cases 1,2,3 for 20-meter depth within 2000 and 4000 seconds. a - Scenario 1; b - Scenario 2.}
    \label{fig:zz_v2_v5}
\end{figure}

Comparing Scenario 1 and 2 (Figures \ref{fig:z_v2}, \ref{fig:dv_dm_v2} and \ref{fig:z_v5}, \ref{fig:dv_dm_v5} respectively), we observe significant differences in the operational demands between the low-speed and high-speed cases for the AUV.
During tests at low speed, as demonstrated in Figures \ref{fig:z_v2} and \ref{fig:dv_dm_v2}, the AUV requires substantial use of actuators and hover tanks. This is primarily due to the need for enhanced stability and precision in maintaining position and orientation in a relatively static state. The control surfaces are utilized extensively to make fine adjustments, which must be highly accurate to counteract waves effectively. 
In contrast, the high-speed test scenario, as illustrated in Figures \ref{fig:z_v5} and \ref{fig:dv_dm_v5}, presents a different situation. 
At higher speeds, the force generated by the control surfaces is greater, leading to more effective maneuverability \cite{martin2022reduced}. Therefore, tail and sail planes are less engaged in constant fine-tuning. Similarly, the role of hover tank is diminished in high-speed operations. As the AUV propels forward, buoyancy control becomes less influential on its stability and direction, with hydrodynamic forces predominantly guiding its motion.

Figure \ref{fig:zz_v2_v5} highlights the impact of augmentation in two distinct scenarios. In the low-speed case, as depicted in Figure \ref{fig:zz_v2_v5}a, the vehicle is predominantly affected by suction forces. Consequently, the augmentation system must exert greater effort to eliminate the steady-state error. Contrastingly, in the high-speed scenario shown in Figure \ref{fig:zz_v2_v5}b, the adaptive controller requires less effort to achieve control objectives due to the reduced influence of these forces at higher velocities. This distinction underscores the varying demands placed on the augmentation system across different operational speeds.


\section{Conclusions}

This paper described a method for controlling the depth and pitch of AUVs using a combination of an LQR controller and \(\mathcal{L}_1\) Adaptive Autopilot. This approach demonstrated substantial performance in control accuracy and stability, particularly in challenging wave conditions near the surface, by filtering out wave disturbances. The study contributed to the field of AUV technology by offering potential solutions for managing dynamic, nonlinear underwater environments. Future work will investigate the altered dynamics of sail plane angles at low speeds, which could further refine control strategies and enhance AUV operational capabilities in various marine settings.


\bibliography{ifacconf}







\end{document}